\RequirePackage{fix-cm}
\documentclass[onecolumn,epjc3]{svjour3}
\RequirePackage{url}
\RequirePackage{graphicx}

\newcommand{\sref}[1]{Sec.~\ref{#1}}
\newcommand{\eref}[1]{Eq.~\ref{#1}}
\newcommand{\fref}[1]{Fig.~\ref{#1}}

\newcommand{\CQG}{{\it Class. Quantum Grav.} }
\newcommand{\GRG}{{\it Gen. Rel. Grav. }}
\newcommand{\JMP}{{\it J. Math. Phys.} }
\newcommand{\PL}{{\it Phys. Lett.} }
\newcommand{\PR}{{\it Phys. Rev.} }
\newcommand{\PRL}{{\it Phys. Rev. Lett.} }
\newcommand{\RPP}{{\it Rep. Prog. Phys.} }

\journalname{Eur. Phys. J. C}

\begin{document}

\title{Newtonian Noise Limit in Atom Interferometers for Gravitational Wave
Detection}

\author{Flavio Vetrano\thanksref{eVetrano,urbino,infn}
            \and
            Andrea Vicer\'e\thanksref{eVicere,urbino,infn}
            }

\thankstext{eVetrano}{flavio.vetrano@uniurb.it}
\thankstext{eVicere}{andrea.vicere@uniurb.it}

\institute{Dipartimento di Scienze di Base e Fondamenti - DiSBeF, Universit\`a degli Studi di Urbino ``Carlo Bo'', I-61029 Urbino, Italy \label{urbino}
               \and
               INFN, Sezione di Firenze, INFN, I-50019 Sesto Fiorentino, Italy \label{infn}
               }

\date{Received: date / Accepted: date}

\maketitle

\begin{abstract}
In this work we study the influence of the newtonian noise on atom
interferometers applied to the detection of gravitational waves, and
we compute the resulting limits to the sensitivity in two different
configurations: a single atom interferometer, or a pair of atom interferometers
operated in a differential configuration. We find that for the instrumental
configurations considered, and operating in the frequency range $[0.1-10]$
Hz, the limits would be comparable to those affecting large scale
optical interferometers.
\end{abstract}

\section{Introduction}

The direct detection of Gravitational Waves is one of the most exciting
challenges of current scientific research. The first generation of
ground-based optical interferometric detectors, including Virgo~\cite{Virgo}
and GEO600~\cite{GEO} in Europe, and the LIGO~\cite{LIGO} interferometers
in USA, achieved design sensitivity and carried out several science
runs, which set interesting upper limits on several classes of astrophysical
sources~\cite{CBClowMass:2012,GRB:2012,Vela:2011,SGWB:2009}. The
construction of a \textquotedblleft{}second generation\textquotedblright{}
of optical interferometers, Advanced LIGO~\cite{aLIGO:2010} and
Virgo~\cite{AdV}, and the new Japanese detector KAGRA~\cite{Kagra},
is well underway; thanks to the implementation of several technical
upgrades, the advanced detectors are expected to come on line with
a sensitivity about ten times better than first generation instruments.
In the meanwhile, the conceptual design of third generation detectors,
like the Einstein Telescope~\cite{ETdetector:2010,ETscience:2012},
has started.

For all these optical ground based detectors the sensitivity
in the low frequency band, below 10 Hz, is ultimately limited by the
so called ``gravity gradient'', or Newtonian Noise (NN)~\cite{Spero:1983,Saulson:1983},
whose source is the direct coupling of the test masses with any mass-density
change in the environment, especially of seismic or atmospheric origin.

Atom interferometers (see~\cite{Berman:1997} for a review) have
been proposed recently as GW detectors~\cite{Vetrano:2004,Chiao:Speliotopoulos:2004,Roura:etal:2006,Delva:etal:2006,Tino:Vetrano:2007,Dimopoulos:2008},
on the basis of previous general ideas~\cite{Borde:1989}.
These instruments promise to be less sensitive to some of the noise sources affecting optical instruments: for instance, being the atoms in free fall, no direct seismic noise should be present. The effect of gravitational waves is a change in the phase accumulated by atoms' wave functions, which can be detected by observing the interference of two atom beams.

However, also the non-radiative gravitational fields of terrestrial origin affect the phase, in a different way as we will show: the question arises then,
if the \textquotedblleft{}low frequency wall\textquotedblright{}
due to NN is relevant also for these new proposed detectors. In this
paper we consider only the NN of seismic origin and we carry out a
detailed calculation of its contribution to the sensitivity curve
of an atom interferometer both in the \textquotedblleft{}single
detector\textquotedblright{} configuration and in the \textquotedblleft{}coupled
differential\textquotedblright{} configuration.

It is worth underlining that this study is motivated by the different way in which gravitational fluctuations couple
to atom interferometers and to optical interferometers, related to the fact that in the first case the test masses are atoms freely traveling across the instrument.
We anticipate our conclusions: the atom interferometers are subject to NN in a degree
similar to optical interferometers, and therefore will require appropriate
technical solutions to overcome this noise limit in the frequency band below
10Hz.

The paper is organized as follows: in~\sref{sec:generalFormulas}
we consider a definite atom interferometer and we compute its response
to a fluctuating gravity field; in~\sref{sec:singleDetector}
we apply the formulas to the case of a single detector, deriving the
limits on sensitivity; finally in~\sref{sec:differentialConfiguration}
we consider two atom interferometers operated in differential configurations.

\section{Newtonian noise of seismic origin in atom interferometers\label{sec:generalFormulas}}

In optical interferometric GW detectors the test masses are suspended
mirrors: a pendular suspension is indeed the best approximation on
Earth for a freely falling test mass. In atom interferometers instead
the role of test masses is played by atoms in free fall, hence our
intent is to determine the influence of the Newtonian coupling to
an external, time-varying mass distribution, on freely falling masses.

Some general considerations are possible: if the effect originates
from seismic noise, it is driven by an external masses displacement
field,
whose linear power spectral density will generally have the form $\tilde{W}\left(\omega\right)\sim\omega^{-2}$,
mediated by a transfer function from the seism to the test masses
motion behaving also as $\omega^{-2}$~\cite{Saulson:1983,Hughes:Thorne:1998,Beccaria:1998},
where $\omega$ is the angular frequency. Therefore the effect on
test masses is expected to be of the form $\theta\left(\omega\right)\Gamma\omega^{-4}$,
hence more relevant at low frequencies,
where $\theta\left(\omega\right)$ is a kind of reduced transfer function,
depending on the detection device, and
$\Gamma$ is a scale factor depending on the model of seismic waves
(it is recognized that the role of main source is played by Rayleigh
surface waves, especially the fundamental mode and few overtones~\cite{Hughes:Thorne:1998,Beccaria:1998}).

To derive the actual expression of $\theta\left(\omega\right)$ for
NN in an atom interferometer, we use the ABCD formalism for matter
waves, described elsewhere in detail~\cite{Tino:Vetrano:2007,Borde:2004}.

Assume that the Hamiltonian of the motion for the atoms is at most quadratic in momentum and position operators
\begin{eqnarray}
H & = & \sum_{n,r=1}^{3}\left[\frac{1}{2M}p_{n}\beta_{nr}(t)p_{r}+\frac{1}{2}p_{n}\alpha_{nr}(t)q_{r}-\frac{1}{2}q_{n}\delta_{nr}(t)p_{r}+\right.\label{eq:Hamiltonian}\\
 &  & \quad\left.-\frac{M}{2}q_{n}\gamma_{nr}(t)q_{r}+f_{n}(t)p_{n}-M g_{n}(t)q_{n}\right]\nonumber 
\end{eqnarray}
where $p_{n(r)}$ and $q_{n(r)}$ are vectors of momentum and position,
respectively, whereas $\alpha,\,\beta,\,\gamma,\,\delta$ are suitable
square matrices (note that $\delta=-\alpha^{T}$, with $T$ indicates
the transposed matrix), and $M$ is the atom rest mass.

The last term in the Hamiltonian represents the response to the local, fluctuating gravitational field $\vec{g}(t)$: in the following, we will consider only the component along the direction of motion of the atoms, as in the paraxial approximation all transverse effects are neglected.
The $\gamma$ term allows to model the response to gravitational waves: in the Fermi gauge, and considering Fourier components, one can show that $\hat{\gamma}=\frac{\omega^2}{2}\hat{h}(\omega)$, where $\hat{h}(\omega)$ is the gravitational wave strain tensor (see for instance~\cite{Tino:Vetrano:2007}).

Consider an
atoms' beam (a Gaussian packet under paraxial approximation~\cite{Tino:Vetrano:2007,Borde:2004,Riou:etal:2008,Borde:2008,Tino:Vetrano:2011})
which is divided and recombined through a sequence of $R$ light-field
beam splitters, supplied by the same laser: from the first beam splitter
to the last one (the output port) we may identify two paths, conventionally
labeled $s$ and $i$. By exploiting the ttt theorem~\cite{Borde:2004}
for the atoms/beam splitter interactions, and the mid-point property
of Gaussian beams~\cite{Borde:2002}, the phase difference at the
output port of the interferometer can be written as:
\begin{equation}
\Delta\phi=\sum_{j=1}^{R}\left[\left(k_{sj}-k_{ij}\right)\frac{q_{sj}+q_{ij}}{2}-\left(\omega_{sj}-\omega_{ij}\right)t_{j}+\left(\theta_{sj}+\theta_{ij}\right)\right]\label{eq:phase}
\end{equation}
where $k_{s(i)j}$ is the momentum transferred to the atoms by the
\emph{j}-th beam splitter along the $s\,\left(i\right)$ arm, $\omega_{s(i)j}$
is the angular frequency of the laser beam and $\theta_{s(i)j}$ is
the phase of the laser beam at the \emph{j}-th interaction, $q_{s(i)j}$
is the distance of j-th interaction point from the laser source; equal
masses are assumed for the atoms along the \emph{s} and \emph{i} paths.
The expression in~\eref{eq:phase} is manifestly gauge-invariant~\cite{Tino:Vetrano:2007,Borde:2004},
and the evolution of the wave packets can be obtained, by means of
the Ehrenfest theorem, from Hamilton's equations for
the vector $\chi(t)$~\cite{Tino:Vetrano:2007,Borde:2004,Tino:Vetrano:2011}
\begin{equation}
\frac{d\chi}{dt}=\left(\begin{array}{c}
\frac{dH}{dp}\\
-\frac{1}{M}\frac{dH}{dq}
\end{array}\right)=\Gamma(t)\cdot\chi(t)+\Phi(t)\label{eq:HamiltonEqs}
\end{equation}
where
\begin{equation}
\chi\equiv\left(\begin{array}{c}
q\\
\frac{p}{M}
\end{array}\right)\,;\quad\Phi(t)\equiv\left(\begin{array}{c}
f(t)\\
g(t)
\end{array}\right)\,;\quad\Gamma(t)\equiv\left(\begin{array}{cc}
\alpha(t) & \beta(t)\\
\gamma(t) & \delta(t)
\end{array}\right)\label{eq:defChiPhiGamma}
\end{equation}
in the form
\begin{equation}
\chi(t)=\left(\begin{array}{cc}
A(t,t_{0}) & B(t,t_{0})\\
C(t,t_{0}) & D(t,t_{0})
\end{array}\right)\cdot\left[\chi\left(t_{0}\right)+\left(\begin{array}{c}
\xi(t,t_{0})\\
\psi(t,t_{0})
\end{array}\right)\right]\label{eq:solChi}
\end{equation}
where
\begin{eqnarray}
\left(\begin{array}{cc}
A(t,t_{0}) & B(t,t_{0})\\
C(t,t_{0}) & D(t,t_{0})
\end{array}\right) & = & \tau\exp\left[\int_{t_{0}}^{t}\Gamma\left(t'\right)dt'\right]\,,\label{eq:timeOrderedExponential}\\
\left(\begin{array}{c}
\xi(t,t_{0})\\
\psi(t,t_{0})
\end{array}\right) & = & \int_{t_{0}}^{t}\left(\begin{array}{cc}
A(t_{0},t') & B(t_{0},t')\\
C(t_{0},t') & D(t_{0},t')
\end{array}\right)\cdot\Phi\left(t'\right)dt'\,;\label{eq:xiPsi}
\end{eqnarray}
here $\tau$ represents the time-ordering operator, and an appropriate
perturbative expansion can be used to evaluate the time-ordered exponential
in~\eref{eq:timeOrderedExponential} \cite{Tino:Vetrano:2007,Borde:2004,Tino:Vetrano:2011}.

\begin{figure}[th]
\centering{}\includegraphics[clip,scale=0.5]{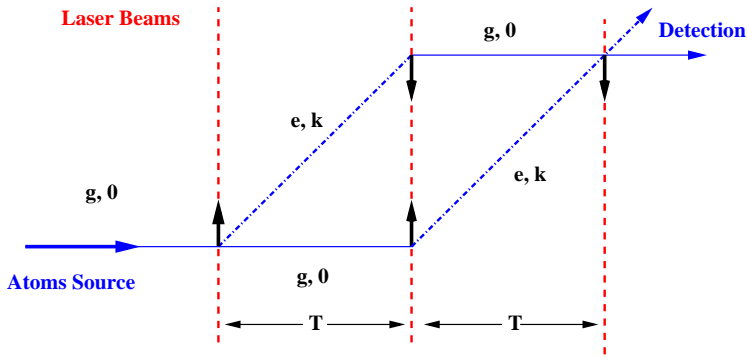}\caption{A simple "Ramsey-Bord\'e atom interferometer with Mach-Zehnder geometry. Continuous horizontal lines, and the slanted dot-dashed lines, represent atom
beams. Vertical dashed lines represent the laser beams; the bold continuous
arrows represent relevant momentum transferred to the atoms; \emph{g}
and \emph{e} mark the ground and excited internal states of the atoms;
\textbf{k} is the transverse momentum in $\hbar$ units. \label{fig:Mach-Zehnder}}
\end{figure}

As a simple reference configuration let us consider a ``Ramsey-Bord\'e''
atom interferometer, with a Mach-Zehnder geometry, as outlined in~\fref{fig:Mach-Zehnder}~\cite{Berman:1997,Tino:Vetrano:2007,Borde:2004}.
In the following, we will also assume that the instrument is crossed
by a plane GW with \textquotedblleft{}+\textquotedblright{} polarization
and amplitude $h$, propagating along the $x_{3}=z$ axis, perpendicular
to the plane of the interferometer; we adopt in the following a description
in Fermi coordinates, which represents the best approximation to the
Laboratory Cartesian system~\cite{Manasse:Misner:1963}.

Assuming the same \textquotedblleft{}stable\textquotedblright{} frequency
for the laser beams and neglecting the steady proper laser phases,
the phase shift formula in~\eref{eq:phase} becomes
\begin{equation}
\Delta\phi=\sum_{j=1}^{4}\left(k_{sj}-k_{ij}\right)\frac{q_{sj}+q_{ij}}{2}\,.
\end{equation}
Let us assume that atoms are subjected only to a fluctuating gravitational
field $g(t)$. Considering~\eref{eq:Hamiltonian},~\eref{eq:HamiltonEqs},~\eref{eq:defChiPhiGamma}~and~\eref{eq:xiPsi}
we have
\begin{eqnarray}
\alpha & = & \delta=\gamma=0\,;\quad\beta=1\,;\quad f(t)=0\,;\quad g(t)\neq0\nonumber \\
A & = & 1\,;\quad B=t-t_{0}\,;\quad C=0\,;\quad D=1\,;
\end{eqnarray}
and we obtain
\begin{equation}
\left(\begin{array}{c}
\xi\left(t,\, t_{0}\right)\\
\psi\left(t,\, t_{0}\right)
\end{array}\right)=\int_{t_{0}}^{t}\left(\begin{array}{c}
t_{0}-t'\\
t'
\end{array}\right)g\left(t'\right)dt'\,.
\end{equation}
We are interested in the low frequency range, where the newtonian
noise is expected to be the limiting factor on account of its $\omega^{-4}$
shape. We will therefore assume that the single atom interferometer
has a linear dimension smaller than the wavelength of seismic surface
waves, which we will assume to set also the coherence length.
Introducing the Fourier transform $\hat{g}\left(\omega\right)$ of the fluctuating
field we can also write
\begin{eqnarray}
\xi\left(t,\, t_{0}\right) & = & \int\frac{d\omega}{2\pi}\hat{g}\left(\omega\right)\left[-\frac{\left(t-t_{0}\right)}{i\omega}e^{i\omega t}-\frac{1}{\omega^{2}}\left(e^{i\omega t}-e^{i\omega t_{0}}\right)\right]\nonumber \\
\psi\left(t,\, t_{0}\right) & = & \int\frac{d\omega}{2\pi}\hat{g}\left(\omega\right)\left[\frac{e^{i\omega t_{0}}}{i\omega}\left(e^{i\omega\left(t-t_{0}\right)}-1\right)\right]\label{eq:xiPsiFreqDom}
\end{eqnarray}
and we assume, in the long wavelength approximation, that $\hat{g}\left(\omega\right)$
is the same at any point of the interferometer. Therefore the solution
of the Hamilton equations~\eref{eq:solChi} can be written as
\begin{eqnarray}
\left(\begin{array}{c}
q(t)\\
\frac{p(t)}{M}
\end{array}\right) & = & \left(\begin{array}{cc}
1 & t-t_{0}\\
0 & 1
\end{array}\right)\cdot\left[\left(\begin{array}{c}
q(t_{0})\\
\frac{p(t_{0})}{M}
\end{array}\right)+\right.\label{eq:qpSol}\\
 &  & \left.\int\frac{d\omega}{2\pi}\hat{g}\left(\omega\right)\left(\begin{array}{c}
-\frac{\left(t-t_{0}\right)}{i\omega}e^{i\omega t}-\frac{1}{\omega^{2}}\left(e^{i\omega t}-e^{i\omega t_{0}}\right)\\
\frac{e^{i\omega t_{0}}}{i\omega}\left(e^{i\omega\left(t-t_{0}\right)}-1\right)
\end{array}\right)\right]\,;\nonumber 
\end{eqnarray}
this expression allows to compute the values of the coordinates and momenta of the atoms at the interaction points with the laser:
by iterating the relation in~\eref{eq:solChi} to the four interaction
points of the interferometer in~\fref{fig:Mach-Zehnder}, setting
$t_{3}=t_{2}$ and defining $T=t_{4}-t_{3}=t_{2}-t_{1}$, we finally
obtain the phase shift at the output port of the interferometer:
\begin{equation}
\Delta\hat{\phi}\left(\omega\right)=kT^{2}e^{i\omega T}\left[\frac{\sin\left(\omega T/2\right)}{\left(\omega T/2\right)}\right]^{2}\hat{g}\left(\omega\right)\,;\label{eq:phaseShift}
\end{equation}
this is the fundamental formula to estimate the effect of the fluctuating
field $\hat{g}$. We recall that $k$ is the unperturbed wave vector
of the laser beam, corresponding to the impulse (in units of the reduced
Planck constant $\hbar$) transferred to the atom at each interaction
point. Note also that in the limit $\omega\rightarrow0$ the expression
in~\eref{eq:phaseShift} corresponds to the well known static
result~\cite{Borde:2002,Borde:2001}.

\section{Newtonian-Noise limit on sensitivity: the single detector case\label{sec:singleDetector}}

In the weak field approximation, to first order in the amplitude $h$
of an impinging gravitational wave, the phase shift at the output
of the interferometer in~\fref{fig:Mach-Zehnder} has been already
obtained in a fully covariant way~\cite{Tino:Vetrano:2007}. Indicating
with $q_{1}$ the unperturbed distance of the first interaction point
from the laser, and with $p_{1}$ the unperturbed momentum of the
atoms, just before the first interaction with the laser beam, we recall
that the Fourier transform of the phase shift, as a function of the
Fourier transformed amplitude $\hat{h}$ of the GW, can be written
as
\begin{eqnarray}
\Delta\hat{\phi}(\omega) & = & \omega\hat{h}(\omega)\frac{T^{2}k}{M}\left(p_{1}+\frac{k\hbar}{2}\right)\times\left[\frac{e^{i\omega T}-e^{2i\omega T}}{\omega T}+i\, e^{i\omega T}\left(\frac{\sin\left(\omega T/2\right)}{\omega T/2}\right)^{2}\right]+\nonumber \\
 &  & +\frac{\omega^{2}\hat{h}\left(\omega\right)}{2}T^{2}kq_{1}\left(\frac{\sin\left(\omega T/2\right)}{\omega T/2}\right)^{2}e^{i\omega T}
 \label{eq:responseToGW}
\end{eqnarray}
in which the proper laser phases have been neglected.

Comparing with the expression of the response to a fluctuating local gravity field~\eref{eq:phaseShift}, we note that the second term of~\eref{eq:responseToGW} corresponds to it, with the substitution $\tilde{g}\rightarrow \frac{q_1}{2}\omega^2\tilde{h}$: however, the overall response to GWs includes also a dynamic term depending on the atom momentum $p_1$ and on the momentum $k$ transferred to the atoms: hence the effects of the local gravitational field and of the gravitational waves are in principle distinguishable.

For a single interferometer with the laser source close to the device, actually the last
term can be neglected and the more relevant one is the term proportional to $p_{1}$,
since we can also generally neglect the recoil term $\frac{k\hbar}{2M}$.
This expression can be directly translated into a relation between
linear power spectral densities (LPSD), that we denote by a tilde,
defined in terms of the two-point correlation functions as 
\begin{equation}
\left<\hat{g}\left(\omega\right)\hat{g}\left(\omega'\right)\right>=2\pi\delta\left(\omega-\omega'\right)\tilde{g}^{2}\left(\omega\right)
\end{equation}
in which the angular brackets represent the statistical average. From~\eref{eq:phaseShift}~and~\eref{eq:responseToGW} we obtain
\begin{eqnarray}
\Delta\tilde{\phi}(\omega) & = & \tilde{h}(\omega)kL\left|\sin\left(\omega T/2\right)\right|\sqrt{1-\frac{2\sin\left(\omega T\right)}{\omega T}+\left[\frac{\sin\left(\omega T/2\right)}{\left(\omega T/2\right)}\right]^{2}}\nonumber \\
\Delta\tilde{\phi}\left(\omega\right) & = & kT^{2}\left[\frac{\sin\left(\omega T/2\right)}{\left(\omega T/2\right)}\right]^{2}\tilde{g}\left(\omega\right)
\end{eqnarray}
where the distance $L=2Tp_{1}/M$ travelled by the atoms in the interferometer of~\fref{fig:Mach-Zehnder} has been introduced;
combining the two equations, we deduce the expression 
\begin{equation}
\tilde{h}_{NN}\left(\omega\right)=\frac{4}{\omega^{2}}\frac{\left|\sin\left(\omega T/2\right)\right|}{\sqrt{1-\frac{2\sin\left(\omega T\right)}{\omega T}+\left[\frac{\sin\left(\omega T/2\right)}{\left(\omega T/2\right)}\right]^{2}}}\frac{\tilde{g}\left(\omega\right)}{L}\label{eq:hNNmz}
\end{equation}
for the equivalent strain $\tilde{h}_{NN}$ induced by the fluctuating field $\tilde{g}\left(\omega\right)$. 

It is useful to discuss here the scale of the $\tilde{g}\mbox{\ensuremath{\left(\omega\right)}}$
LPSD, referring to typical values measured at the site of the Virgo
interferometers; we recall indeed that we are considering the effect
of an external fluctuating gravity field on freely falling test masses,
which is the same situation experienced by the test masses of optical
interferometers~\cite{Saulson:1983,Hughes:Thorne:1998,Beccaria:1998};
even though the detailed shape of the NN affecting a instrument like
Virgo depends on the model for the seismic sources and the superficial
Earth layers, similar results are obtained in different cases, which
can be summarized as follows
\begin{equation}
\tilde{h}_{NN}\left(\omega\right)=\frac{\sqrt{4}\tilde{X}\left(\omega\right)}{L_{V}}\simeq\frac{1.2\times10^{-9}}{\omega^{2}}\tilde{x}_{seism}\left(\omega\right)\times\frac{Hz^{2}}{m}\label{eq:virgoNewtonianNoise}
\end{equation}
where $L_{V}=3000$m is the length of Virgo arms, $\tilde{X}\left(\omega\right)$
is the displacement LPSD for a single suspended mirror, and $\tilde{x}_{seism}\left(\omega\right)$
is the measured LPSD of the ground seism~\cite{VirgoSensitivity};
the factor $\sqrt{4}$ takes into account that in Virgo the noise
due to the four end-station mirrors adds in quadrature.

Considering the relation between the mirror motion and its acceleration, due to
the fluctuating gravitational field, $\tilde{g}\left(\omega\right)=\omega^{2}\tilde{X}\left(\omega\right)$,
we obtain
\begin{equation}
\frac{\tilde{g}\left(\omega\right)}{L}=\frac{\omega^{2}L_{V}}{2\, L}\frac{\sqrt{4}\tilde{X}\left(\omega\right)}{L_{V}}\simeq6\times10^{-10}\frac{L_{V}}{L}\tilde{x}_{seism}\left(\omega\right)\times\frac{\mbox{H\ensuremath{z^{2}}}}{\mbox{m}}\,;
\end{equation}
we further assume that the seismic noise measured at the Virgo site
is well approximated by~\cite{SeismicNoiseVirgo}
\begin{equation}
\tilde{x}_{seism}\left(\omega\right)\simeq\frac{10^{-7}}{\left[\omega/\left(2\pi\mbox{Hz}\right)\right]^{2}}\mbox{m}\,\mbox{\mbox{Hz}}^{-1/2}\ ;\label{eq:seismicNoiseVirgo}
\end{equation}
Following~\cite{Tino:Vetrano:2011}, let us assume very ambitious parameters for the single Ramsey-Bord\'e atom interferometer: a length $L\sim~200$m, which could result in interesting sensitivities to gravitational waves, and a time of flight $T=0.4$s, in order to have not too small a bandwidth; obviously the choice implies atom speeds of the order of 250 m/s, and we underline that such choices are probably beyond the limits of current technologies. Anyway, we obtain
\begin{equation}
\frac{\tilde{g}\left(\omega\right)}{L}\sim\frac{10^{-16}}{\left[\omega/\left(2\pi\mbox{Hz}\right)\right]^{2}}\mbox{Hz}^{2}\,.\label{eq:gNoiseVirgo}
\end{equation}
as an estimate of the scale of the fluctuating gravitational field seen by the atom interferometer.

\begin{figure}[th]
\begin{centering}
\includegraphics[clip,scale=0.5]{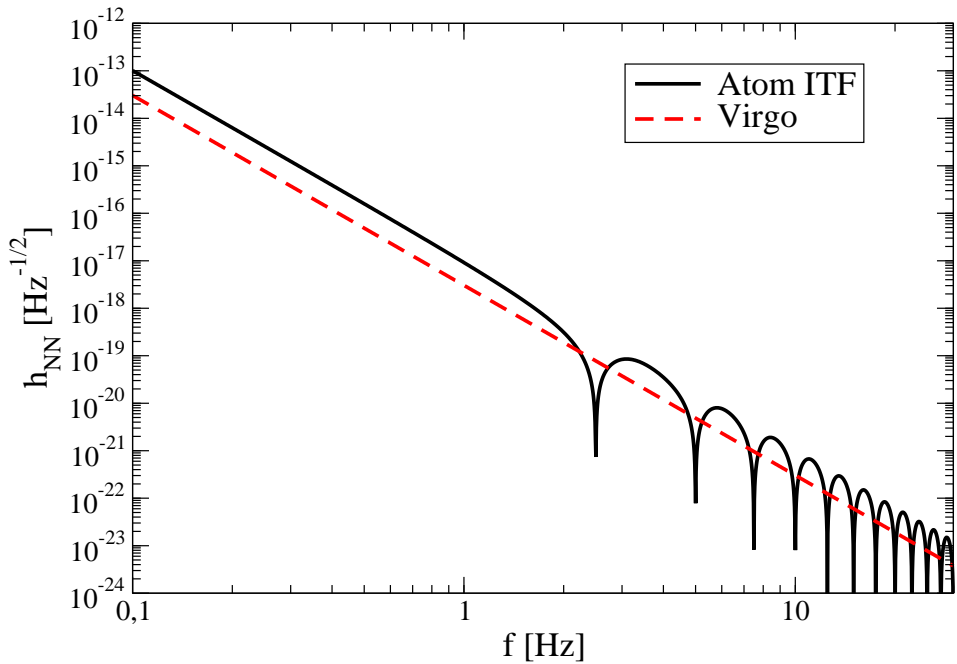}
\par\end{centering}
\caption{\label{fig:hNNsimple}The solid curve represents the effect of the
gravity gradient noise on a single atom interferometer, with the expected
$\omega^{-4}$ behavior, and zeroes corresponding to frequencies at
which the instrument is insensitive both to the gravity gradient fluctuations
and to gravitational waves. For comparison, the dashed curve represents
the model newtonian noise effect on the Virgo interferometer.}
\end{figure}

To appreciate the result, we show in~\fref{fig:hNNsimple} a example of the newtonian noise
of~\eref{eq:hNNmz} assuming the expression in~\eref{eq:gNoiseVirgo}
for the LPSD of the fluctuating gravitational field; in the same figure we plot, for comparison,
the corresponding newtonian noise for the Virgo detector\footnote{It should be underlined that in this low frequency band, below $10$ Hz, the actual noise of Virgo is dominated by other noise sources, most notably by the direct seismic noise and by the thermal noise, not to mention other technical noises.}.

The zeroes represent frequencies at which the atom interferometer is insensitive both to the gravity gradient noise and to GW; note that the one shown is not a complete noise budget, to which other noises would contribute, particularly the atom shot noise which would exhibit peaks at those frequencies, not differently from an optical interferometer in a Michelson configuration and without Fabry-Perot cavities.

Apart this specific feature, the comparison with a large optical interferometer shows a similar behavior as a function of the frequency, with a different noise scale dictated by the different linear dimensions of the instruments. We underline that for this type of atom interferometer, it could be unrealistic to increase the linear size $L$ even further: to this end, a differential configuration appears more promising. 

\section{Two detectors operated in a differential configuration\label{sec:differentialConfiguration}}

Let us now consider the second term in~\eref{eq:responseToGW},
proportional to the position $q_{1}$. This term, already introduced
in a different context~\cite{Borde:etal:1983}, is a sort of ``clock''
term which takes into account the influence of the GW on the laser
beam, along its path from the source to a well defined physical point.
Its role was discussed in recent papers~\cite{Mueller:2009,hoensee:2011,hogan:2011}
and the most relevant new property is the introduction of $q_{1}$
(path of laser beam) in place of $L$ (path of atom beam); so, in
order to improve the sensitivity, enlarging $q_{1}$ seems in principle
easier than enlarging $L$. 

This solution requires measuring the distance from the laser, and carries additional requirements
on the coherence and stability of the laser beam, while maintaining it at a sufficient power density:
it is therefore premature to draw too optimistic conclusions about the practicality of the configuration.
However, the idea of adopting a two-interferometers differential configuration~\cite{Dimopoulos:2008}
appears very appealing in order to render the system independent from the laser position, and may furthermore
yield a good common-modes rejection.

Under the hypothesis of a common laser source for two identical Mach-Zehnder
atom interferometers in differential configuration, for which the
relative distance $D$ satisfies the condition $\omega D/c\ll1$ (with
$c$ the speed of light in vacuum), from~\eref{eq:responseToGW}
the overall difference between the two partial phase differences at
the output ports can be formally obtained as
\begin{equation}
\Delta\hat{\phi}\left(\omega\right)=2k\, D\sin^{2}\left(\omega T/2\right)e^{i\omega T}\hat{h}\left(\omega\right)\label{eq:secPhiDiff}
\end{equation}
where $D\equiv q_{1}^{II}-q_{1}^{I}$ as anticipated.
Considering also~\eref{eq:phaseShift} we obtain for the differential configuration
\begin{equation}
\hat{h}_{NN}\left(\omega\right)=\frac{2}{\omega^{2}D}\left[\hat{g}_{2}\left(\omega\right)-\hat{g}_{1}\left(\omega\right)\right]\,,\label{eq:hNNdiffConf}
\end{equation}
where the difference in the right hand side requires some discussion.
In a given frequency band, if the two fluctuating gravity fields $\hat{g}_{1,2}$
act upon sufficiently distant atom interferometers, they will be uncorrelated,
and we will obtain for the LPSD simply a sum in quadrature
\begin{equation}
\tilde{h}_{NN}\left(\omega\right)=\frac{2}{\omega^{2}D}\sqrt{\tilde{g}_{1}^{2}\left(\omega\right)+\tilde{g}_{2}^{2}\left(\omega\right)}
\end{equation}
displaying no conceptual difference with respect to the limits obtained
for optical interferometers with long arms~\cite{VirgoSensitivity}.
Considering instead a low-frequency, long-wavelength approximation,
it may be appealing the situation in which, even with two separated
interferometers, the residual correlation leads to a partial noise
cancellation in~\eref{eq:hNNdiffConf}.

\begin{figure}[th]
\centering{}\includegraphics[scale=0.6]{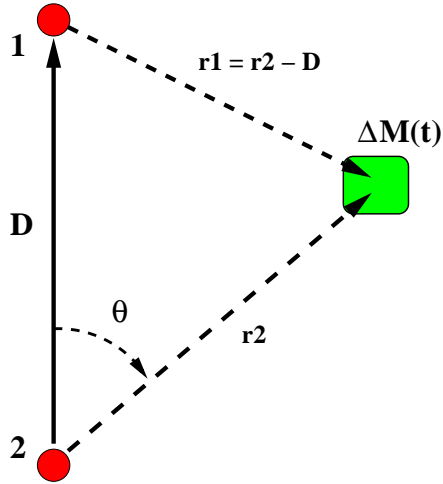}\caption{\label{fig:diffGeometry}Geometry of the detector: atom interferometers are located at positions 1 and 2, and a fluctuating mass element is
assumed at a location $\vec{r}_{2}$ in a frame having position 2
as the origin, and a $\hat{z}$ axis parallel to $\vec{D}$.}
\end{figure}

We recall that the signals $\hat{g}_{1,2}\left(\omega\right)$ are
assumed to be stochastic acceleration fields in positions 1 and 2,
projected along the direction specified by the segment $\vec{D}$
as in~\fref{fig:diffGeometry}.

We further assume to model the stochastic noise in the simplest possible
way, namely as due to uncorrelated fluctuations in the density of
the material surrounding the detector~\cite{Saulson:1983}. In other
words a density fluctuation $\Delta M\left(t\right)$ will contribute
to the acceleration field in points 1 and 2 as
\begin{eqnarray}
\vec{g}_{2}(t) & = & \frac{G\Delta M(t)}{r_{2}^{2}}\hat{r}_{2}=\frac{G\Delta M(t)}{r_{2}^{3}}\vec{r}_{2}\\
\vec{g}_{1}(t) & = & \frac{G\Delta M(t)}{r_{1}^{2}}\hat{r}_{1}=\frac{G\Delta M(t)}{\left|r_{2}-D\right|^{3}}\left(\vec{r}_{2}-\vec{D}\right)
\end{eqnarray}
Considering only the component acting along the direction separating
the two points 1 and 2, we obtain
\begin{eqnarray}
g_{2}(t) & = & \frac{G\Delta M(t)}{r^{2}}\cos\left(\theta\right)\nonumber \\
g_{1}(t) & = & \frac{G\Delta M(t)}{\left[r^{2}+D^{2}-2rD\cos\left(\theta\right)\right]^{3/2}}\left[r\cos\left(\theta\right)-D\right]
\end{eqnarray}
as the contribution to the fluctuation of the acceleration field due to a single mass element.
To obtain the total fluctuation, we need now to sum over the space.

We first assume for simplicity that the space around the two stations with atom interferometers can be considered homogeneous:
this could be the case for instance if the instrumentation is placed in a deep mine, at a depth much larger than $D$.
We are therefore interested in the quantity
\begin{eqnarray}
\hat{h}_{NN}\left(\omega,\,\vec{r}\right) & = & \frac{2}{\omega^{2}D}\left[\hat{g}_{2}\left(\omega\right)-\hat{g}_{1}\left(\omega\right)\right]\\
& = & \frac{2G\Delta M\left(\omega,\,\vec{r}\right)}{\omega^{2}D}\left\{ \frac{\cos\left(\theta\right)}{r^{2}}-\frac{r\cos\left(\theta\right)-D}{\left[r^{2}+D^{2}-2rD\cos\left(\theta\right)\right]^{3/2}}\right\} \nonumber 
\end{eqnarray}
which should be summed over the volume. It is convenient to evaluate the spectral density
\begin{eqnarray}
\left\langle h_{NN}\left(\omega\right)h_{NN}\left(\omega'\right)\right\rangle  & \equiv & 2\pi\delta\left(\omega-\omega'\right)\tilde{h}_{NN}^{2}\left(\omega\right)\\
 & = & \sum_{\vec{r},\,\vec{r'}}\left\langle \Delta h_{NN}\left(\omega,\,\vec{r}\right)\Delta h_{NN}\left(\omega',\,\vec{r'}\right)\right\rangle \,;\nonumber 
\end{eqnarray}
where, following again Saulson~\cite{Saulson:1983}, we assume the
sum to be extended over volume elements of linear size $\lambda/2,$
with $\Delta M$ fluctuating coherently inside these regions, and
totally uncorrelated otherwise: 
\begin{equation}
\left\langle \Delta M\left(\omega,\,\vec{r}\right)\Delta M\left(\omega',\,\vec{r}'\right)\right\rangle =2\pi\delta\left(\omega-\omega'\right)\Delta\tilde{M}^{2}\left(\omega,\,\vec{r}\right)\delta_{\vec{r},\vec{r'}}\,.
\end{equation}
We obtain therefore
\begin{equation}
\tilde{h}_{NN}^{2}\left(\omega\right)=\frac{4G^{2}}{\omega^{4}D^{2}}\sum_{\vec{r}}\Delta\tilde{M}^{2}\left(\omega,\,\vec{r}\right)\left\{ \frac{\cos\left(\theta\right)}{r^{2}}-\frac{r\cos\left(\theta\right)-D}{\left[r^{2}+D^{2}-2rD\cos\left(\theta\right)\right]^{3/2}}\right\} ^{2}\,.
\end{equation}
If we additionally assume that the mass fluctuations do not depend on $\vec{r}$, we can further simplify, obtaining
\begin{eqnarray}
\tilde{h}_{NN}^{2}\left(\omega\right) & = & \frac{4G^{2}\Delta\tilde{M}^{2}\left(\omega\right)}{\omega^{4}D^{2}}\sum_{\vec{r}}\left\{ \frac{\cos\left(\theta\right)}{r^{2}}-\frac{r\cos\left(\theta\right)-D}{\left[r^{2}+D^{2}-2rD\cos\left(\theta\right)\right]^{3/2}}\right\} ^{2}\\
& = & \frac{4G^{2}\Delta\tilde{M}^{2}\left(\omega\right)}{\omega^{4}D^{2}}\left(\frac{2}{\lambda}\right)^{3}\int\left\{ \frac{\cos\left(\theta\right)}{r^{2}}-\frac{r\cos\left(\theta\right)-D}{\left[r^{2}+D^{2}-2rD\cos\left(\theta\right)\right]^{3/2}}\right\} ^{2}r^{2}dr\, d\cos\theta\, d\phi\,,\nonumber 
\end{eqnarray}
where we have approximated the sum with an integral, normalizing by
the volume element of the coherent region $\left(\lambda/2\right)^{3}$.
If we were to retain only the first term, we would obtain the same
result as in~\cite{Saulson:1983}, corrected for a factor 2 which is wrong in the original paper.
The integration over the angular functions is directly carried out, resulting in a lengthy expression:
\begin{eqnarray}
\tilde{h}_{NN}^{2}\left(\omega\right) & = & \frac{64\pi G^{2}\Delta\tilde{M}^{2}\left(\omega\right)}{\omega^{4}D^{2}\lambda^{3}}\cdot H\left(D,\,\lambda\right)\label{eq:hNNgeneral}\\
H & =\int & \frac{r\left\{ 4\left[8(D-r)^{2}(D+r)^{2}+3Dr\left(3D^{2}-r^{2}\right)\right]-3\left(D^{2}-r^{2}\right)^{2}\ln\frac{(D-r)^{2}}{\left(D+r\right)^{2}}\right\} }{24D^{3}(D-r)^{2}(D+r)^{2}}dr+\nonumber \\
 & + & \int\frac{2\left(D^{3}+2r^{3}\right)\left(D-r\right)}{3D^{3}r^{2}|D-r|}dr\nonumber 
\end{eqnarray}
which, as expected, displays double poles in $r=0$ and in $r=D$.

Both divergences are artefacts, which should be regulated introducing cutoffs $r\geq\frac{\lambda}{4}$ and at $\left|r-D\right|\geq\frac{\lambda}{4}$. However, it is now necessary to distinguish two cases

\paragraph{Short wavelength}

If the distance $D\gg\lambda$, then the integral over $r$ gives
\begin{equation}
H\left(D,\,\lambda\right)=\frac{14}{3\lambda}+O\left(\frac{\lambda}{D^{2}}\ln\frac{\lambda}{D}\right)
\end{equation}
and we obtain
\begin{equation}
\tilde{h}_{NN(sw)}^{2}\left(\omega\right)\simeq\frac{896\,\pi G^{2}\Delta\tilde{M}^{2}\left(\omega\right)}{3\,\omega^{4}D^{2}\lambda^{4}}\,\cdot
\end{equation}

\paragraph{Long wavelength}

In the long wavelength approximation the integral in~\eref{eq:hNNgeneral}
can be carried out assuming $r\geq\frac{\lambda}{4}\gg D$, obtaining
\begin{equation}
H\left(D,\,\lambda\right)=\frac{512\, D^{2}}{15\,\lambda^{3}}+O\left(\frac{D^{4}}{\lambda^{5}}\right)
\end{equation}
hence
\begin{equation}
\tilde{h}_{NN(lw)}^{2}\left(\omega\right)\simeq\frac{32768\,\pi G^{2}\Delta\tilde{M}^{2}\left(\omega\right)}{15\,\omega^{4}\lambda^{6}}\,;
\end{equation}
it seems at first surprising that the dependence on $D$ cancels out in the long wavelength approximation,
whereas one could have expected to retain a dependence,
which could lead to zero the noise in the $D\rightarrow 0$ limit case.
However, we are actually in a situation in which the instrument is
sensitive to the gradient of the gravity acceleration (see~\eref{eq:hNNdiffConf}),
and therefore, barring other sources of noise, the sensitivity is
independent on the baseline $D$. 

We can now use Eq.~12 of~\cite{Beccaria:1998} to relate the mass fluctuations with the measured seism
\begin{equation}
\Delta\tilde{M}^{2}\left(\omega\right)=\frac{1}{16}\lambda^{6}\rho_{0}^{2}\left(\frac{\pi}{\lambda}\right)^{2}\tilde{x}_{seism}^{2}\left(\omega\right)
\end{equation}
where $\rho_{0}$ is the density of the medium. We finally obtain
\begin{eqnarray}
\tilde{h}_{NN(sw)}\left(\omega\right) & \simeq & \frac{2\pi\sqrt{14\pi}G\rho_{0}}{\sqrt{3}\,\omega^{2}D}\tilde{x}_{seism}\left(\omega\right)\label{eq:hNNsw}\\
\tilde{h}_{NN(lw)}\left(\omega\right) & \simeq & \frac{16\sqrt{2\pi}G\rho_{0}}{\sqrt{15}\,\omega c_{L}}\tilde{x}_{seism}\left(\omega\right)\label{eq:hNNlw}
\end{eqnarray}
where we have used the relation $\lambda\omega=2\pi c_{L}$, with $c_{L}$ the speed of longitudinal seismic waves.

Comparing with~\eref{eq:virgoNewtonianNoise} for the gravity gradient noise affecting the Virgo interferometer, we see that in the short wavelength limit, represented by~\eref{eq:hNNsw}, the frequency dependence (as expected) is the same. Instead, in the long wavelength limit~\eref{eq:hNNlw}, the NN affecting the atom interferometer has a slower growth for $\omega\rightarrow0$, reflecting the presence of correlated noise at the two stations, that partially cancels out in~\eref{eq:hNNdiffConf}.

We underline that this cancellation is not specific of a dual atomic interferometer: the same effect would occur in optical interferometers like Virgo, for shorter baselines. However, in optical interferometers long baselines are motivated by the need to reject the mirror position noise, which scales inversely with the distance: in atom interferometers some position noises, like the thermal noise, are instead expected to be absent, hence the baseline could be shorter.

In order to assess the significance of the cancellation effect, we choose favorable, yet realistic parameters: for the medium surrounding our hypothetical instrument, we assume a large $c_L = 5000$~m/s, characteristic of compact rock, and a density $\rho_{0}\simeq2.7\times10^{3}\mbox{kg}\,\mbox{m}^{-3}$, a typical value for the continental crust; we also assume, on the basis of measurement taken in underground environments (for instance in the Kamioka mine~\cite{kamioka:2003} which will host KAGRA) a seismic noise $\tilde{x}_{seism}$ 10 times lower than the one measured at the Virgo site~(\eref{eq:seismicNoiseVirgo}).

We also assume to build a relatively large instrument, taking for the distance between the atom interferometers a value $D\simeq~1$km as proposed in~\cite{dimopoulos:2009}: we obtain
\begin{eqnarray}
\tilde{h}_{NN(sw)}\left(\omega\right) & \simeq & \frac{10^{-18}}{\left[\omega / \left(2\pi\mbox{Hz}\right)\right]^4} \mbox{Hz}^{-1/2}\qquad\frac{\omega}{2\pi}\gg\frac{c_{L}}{D}\simeq 5\mbox{Hz}\\
\tilde{h}_{NN(lw)}\left(\omega\right) & \simeq & \frac{6\times 10^{-19}}{\left[\omega / \left(2\pi\mbox{Hz}\right)\right]^3} \mbox{Hz}^{-1/2}\qquad\frac{\omega}{2\pi}\ll\frac{c_{L}}{4 D}\simeq 1.25 \mbox{Hz}\,.
\end{eqnarray}
\begin{figure}[th]
\begin{centering}
\includegraphics[clip,scale=0.45]{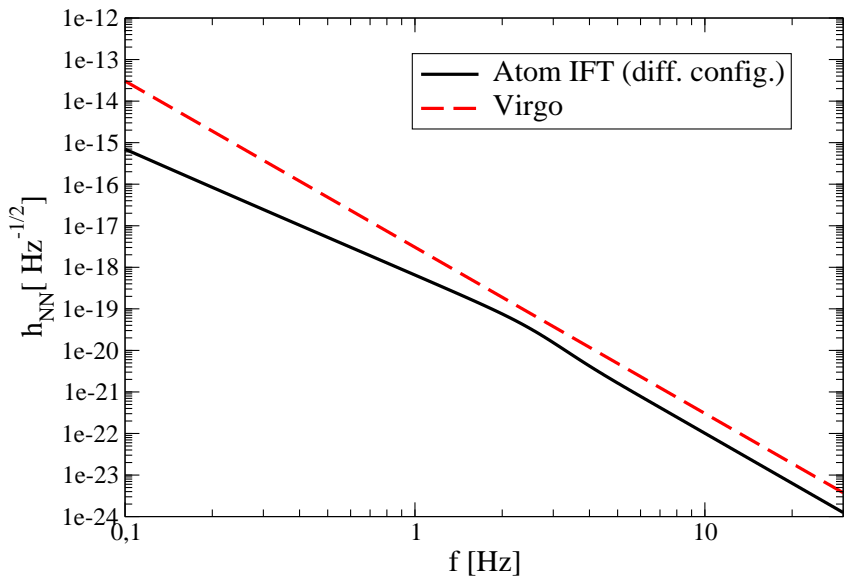}\caption{\label{fig:comparisonNNvirgoAtom}Comparison of models of the Newtonian Noise as seen by the Virgo interferometer (dashed line) or by
an hypothetical pair of atom interferometers operated in differential
configuration (continuous line).
Above a few Hz, the two curves run parallel, at different scales because of the different seismic noise (10 times lower for the hypothetical underground atom interferometers), and the different baseline of the two instruments (3km for the length of Virgo arms, 1km for the distance between the atom interferometers).
At lower frequencies, thanks to its shorter baseline, the dual atom interferometer displays a different slope thanks to the cancellation effect.}
\par\end{centering}
\end{figure}
The resulting limit to the atom interferometer sensitivity is displayed in~\fref{fig:comparisonNNvirgoAtom}, over a frequency range which runs from the long to the short wavelength regimes; for comparison we display also the NN affecting the Virgo instrument; in the high frequency regime, the two curves differ just by a small scale, reflecting the different size of the instruments and the lower seismic noise anticipated for an underground atom interferometer.
In the low frequency regime the residual correlation of the newtonian noise which affects the two atom interferometers, thanks to the shorter baseline, contributes to a milder growth as $\omega\rightarrow0$, and therefore leads to a sizable, though not dramatic, reduction of the noise over the Virgo case.

\section{Conclusions}

In this work we have evaluated the effect of fluctuations of the gravity
field on the sensitivity of atom interferometers, thus providing an
estimate of the so-called newtonian (or gravity gradient) noise for
this kind of instruments.

We have seen that a mid-scale atom interferometer, with a baseline $L\sim200$m, is subject to a noise essentially equivalent to the one affecting a large scale optical interferometer, as Virgo.

We have also found that operating two small-scale atom interferometers, linked by a laser, at a larger distance $D\sim1$ km, in differential configuration, as proposed for instance in~\cite{dimopoulos:2009}, there is an advantage at low frequency thanks to the residual newtonian noise correlation and the resulting partial cancellation.
However, the noise reduction is not dramatic and the newtonian limit remains very significant: it is worth reminding that in order to detect a binary neutron star inspiral (say, at $z\sim 1$) sensitivities better than $10^{-22}$ would be required at $1$ Hz; even for larger systems, say a $1000 M_\odot$ binary black-hole coalescence, sensitivities of the order of $10^{-20}$ should be achieved, as discussed for instance in~\cite{decigo:2008}.

We conclude that, similarly to what is foreseen for future optical interferometers~\cite{ETdetector:2010}, operating successfully atom interferometers in the $[0.1,\,10]$ Hz frequency window will require mitigating the gravity gradient noise; not just by choosing very quiet, underground sites, but also devising clever noise subtraction strategies.

We acknowledge that this study has a limitation in the model for the gravity fluctuations, which is approximate; however, as it has been the case for similar studies carried out for optical interferometers~\cite{Beccaria:1998,Hughes:Thorne:1998},
we believe that the use of more refined models will change the numerical results only by small factors, which would not alter our conclusions.

\end{document}